\documentclass[pra,twocolumn, superscriptaddress,groupedaddress,nofootinbib]{revtex4}  
\usepackage{graphicx}
\usepackage{epstopdf}
\usepackage{amssymb}
\usepackage{amsmath}
\usepackage{xcolor}
\usepackage{amsfonts}
\usepackage{subfigure}
\usepackage{bm}
\usepackage{footnote}
\usepackage{lipsum}
\begin{document}

\title{Realization of a high power optical trapping setup free from thermal lensing effects}

\author{C. Simonelli}
\affiliation{Istituto Nazionale di Ottica del Consiglio Nazionale delle Ricerche (INO-CNR), 50019 Sesto Fiorentino, Italy}
\affiliation{LENS and Dipartimento di Fisica e Astronomia, Universit\`{a} di Firenze, 50019 Sesto Fiorentino, Italy}

\author{E. Neri}
\affiliation{Istituto Nazionale di Ottica del Consiglio Nazionale delle Ricerche (INO-CNR), 50019 Sesto Fiorentino, Italy}
\affiliation{LENS and Dipartimento di Fisica e Astronomia, Universit\`{a} di Firenze, 50019 Sesto Fiorentino, Italy}

\author{A. Ciamei}
\affiliation{Istituto Nazionale di Ottica del Consiglio Nazionale delle Ricerche (INO-CNR), 50019 Sesto Fiorentino, Italy}
\affiliation{LENS and Dipartimento di Fisica e Astronomia, Universit\`{a} di Firenze, 50019 Sesto Fiorentino, Italy}

\author{I. Goti}
\affiliation{LENS and Dipartimento di Fisica e Astronomia, Universit\`{a} di Firenze, 50019 Sesto Fiorentino, Italy}

\author{M. Inguscio}
\affiliation{Istituto Nazionale di Ottica del Consiglio Nazionale delle Ricerche (INO-CNR), 50019 Sesto Fiorentino, Italy}
\affiliation{LENS and Dipartimento di Fisica e Astronomia, Universit\`{a} di Firenze, 50019 Sesto Fiorentino, Italy}

\author{A. Trenkwalder}
\affiliation{Istituto Nazionale di Ottica del Consiglio Nazionale delle Ricerche (INO-CNR), 50019 Sesto Fiorentino, Italy}
\affiliation{LENS and Dipartimento di Fisica e Astronomia, Universit\`{a} di Firenze, 50019 Sesto Fiorentino, Italy}

\author{M. Zaccanti}
\affiliation{Istituto Nazionale di Ottica del Consiglio Nazionale delle Ricerche (INO-CNR), 50019 Sesto Fiorentino, Italy}
\affiliation{LENS and Dipartimento di Fisica e Astronomia, Universit\`{a} di Firenze, 50019 Sesto Fiorentino, Italy}





\begin{abstract}
Transmission of high power laser beams through partially absorbing materials modifies the light propagation via a thermally-induced effect known as thermal lensing. This may cause changes in the beam waist position and degrade the beam quality. Here we characterize the effect of thermal lensing associated with the different elements typically employed in an optical trapping setup for cold atoms experiments. We find that the only relevant thermal lens is represented by the $TeO_2$ crystal of the acousto-optic modulator exploited to adjust the laser power on the atomic sample. We then devise a simple and totally passive scheme that enables to realize an inexpensive optical trapping apparatus essentially free from thermal lensing effects.
\end{abstract}

\maketitle

\date{\today}


\section{Introduction}
The precise focusing of a high power laser beam on a target sample is highly relevant both for fundamental science and for a variety of industrial and medical applications: from the realization of optical tweezers \cite{Neuman_2004} and traps \cite{Grimm_2000} for atoms and molecules, to the exploitation of high power laser sources for cutting, welding, drilling and surface treatment of various materials, to laser-based surgery and ophtalmology.
Quite generally, many applications require the optical power to be controllably tuned, e.g. to enable evaporative cooling of atomic gases in dipole traps, or to avoid undesired damage of the illuminated sample. 
In combination with a high level of optical power, this makes such applications of laser technology not immune from the so-called thermal lensing, or thermal blooming, effect \cite{Gordon_1965,Sparks_1971, Bendow_1973, Sheldon_1982}.
Such a phenomenon arises from the fact that both the substrate and coating of any element composing an optical setup unavoidably absorb part of the incident light. As a consequence, the non-uniform intensity profile of the impinging beam acts as an inhomogeneous heat source for the optical material. Given that the index of refraction inherently features some temperature dependence, the illuminated optical component acts like a lens on the transmitted beam \cite{Gordon_1965,Beausoleil_2003}, making both the size and the location of the beam waist time- and intensity-dependent quantities.
Although thermal lensing effects can be in some cases mitigated by exploiting materials with low absorption coefficients at the laser wavelength of interest, any optical component has inherently an associated thermal lens \cite{Bogan_2015}, which may cause relevant modifications of the beam properties, especially for those instances where stable positioning of the waist is requested at the micro-scale.

In the context of cold gases experiments, high power optical dipole traps (ODT) are routinely employed to confine and manipulate samples of single atomic species or of binary mixtures that cannot be efficiently cooled within magnetic potentials.
Celebrated examples are the case of lithium atoms, see e.g. Refs. \cite{Ohara_2002,Jochim_2003,Burchianti_2014}, and of lithium-potassium mixtures \cite{Wille_2008,Spiegelhalder_2010}: there, an all optical approach is extremely convenient, as it can be employed in combination with external magnetic fields that enable the controlled tuning of the interactions via the Feshbach resonance phenomenon \cite{Chin_2010}. On the other hand, laser sources, generally in the near infrared wavelength regime, delivering powers up to a few hundreds of Watts are unavoidably required to ensure a large trapping volume and trap depths sufficiently high to confine laser-cooled atomic samples delivered by standard magneto-optical traps (or optical molasses) at few hundreds (tens) of $\mu$K.
While thermal lensing does not prevent to reach high efficiencies in confining and manipulating single species within monochromatic traps, it may become a severe limitation in experiments where heteronuclear mixtures or bichromatic potentials are employed, see e.g. Refs. \cite{Onofrio_2002,Spiegelhalder_2010,Tassy_2010,Hansen_2013,Vaidya_2015}.
In the former case, owing to the different polarizabilities of the two atomic species, thermal lensing may induce out-of-phase sloshing of the two clouds within the trap, hence reducing the efficiency of the evaporative and sympathetic cooling stages.
In the latter case, in which the optical potential is realized by superimposing waists of laser beams at different wavelengths, thermal effects may result in an uncontrolled variation of the overall trapping landscape, given that absorption might strongly vary with the frequency of the laser source.
As a consequence, devising schemes to limit, and possibly cancel, thermal lensing effects might significantly increase the performances of cold gases machines based on all-optical approaches.

In this paper we provide a simple and inexpensive strategy to realize a deep dipole trap immune from thermal lensing. This is based on a completely passive setup realized with a 300 Watt laser source at 1070 nm and standard optical elements.
First, we characterize the power of the thermal lens associated with each optical component (lenses, windows, acousto-optic modulator) generally employed within an optical trapping setup. From such a study we conclude that: (i) fused silica lenses and windows with standard anti-reflection coating can be safely used up to powers of several hundreds of Watts, yielding little or no difference with respect to much more expensive elements, such as those based on $Suprasil$\textsuperscript\textregistered substrates; (ii) the only significant thermal lens in the setup is provided by the $TeO_2$ crystal of the acousto-optic modulator (AOM), that represents a typical option to enable the active tuning and control of the laser power on the atomic sample.
Second, we devise, implement and successfully test an optical scheme that allows to precisely cancel the effect of the AOM thermal lens, simply by adjusting the crystal position relative to a focus within the optical path.
We anticipate that, although the present work is primarily targeted to the optical trapping of cold atomic clouds, our study might be straightforwardly extended to any other setup which requires to position the waist of high power lasers on a target sample with a few micron accuracy.              

This article is organized as it follows: Section 2 provides a basic theoretical background to the thermal lensing phenomenon. Section 3 presents a characterization of the thermal lenses associated with the various optical elements employed within a typical optical trapping setup for cold atoms experiments. Finally, Section 4 describes the simple optical scheme we devised to get rid of thermal lensing effects, and the characterization of the resulting ODT beam.

\section{Theoretical background}

\begin{figure}[t!]
\begin{center}
\includegraphics[width=7cm]{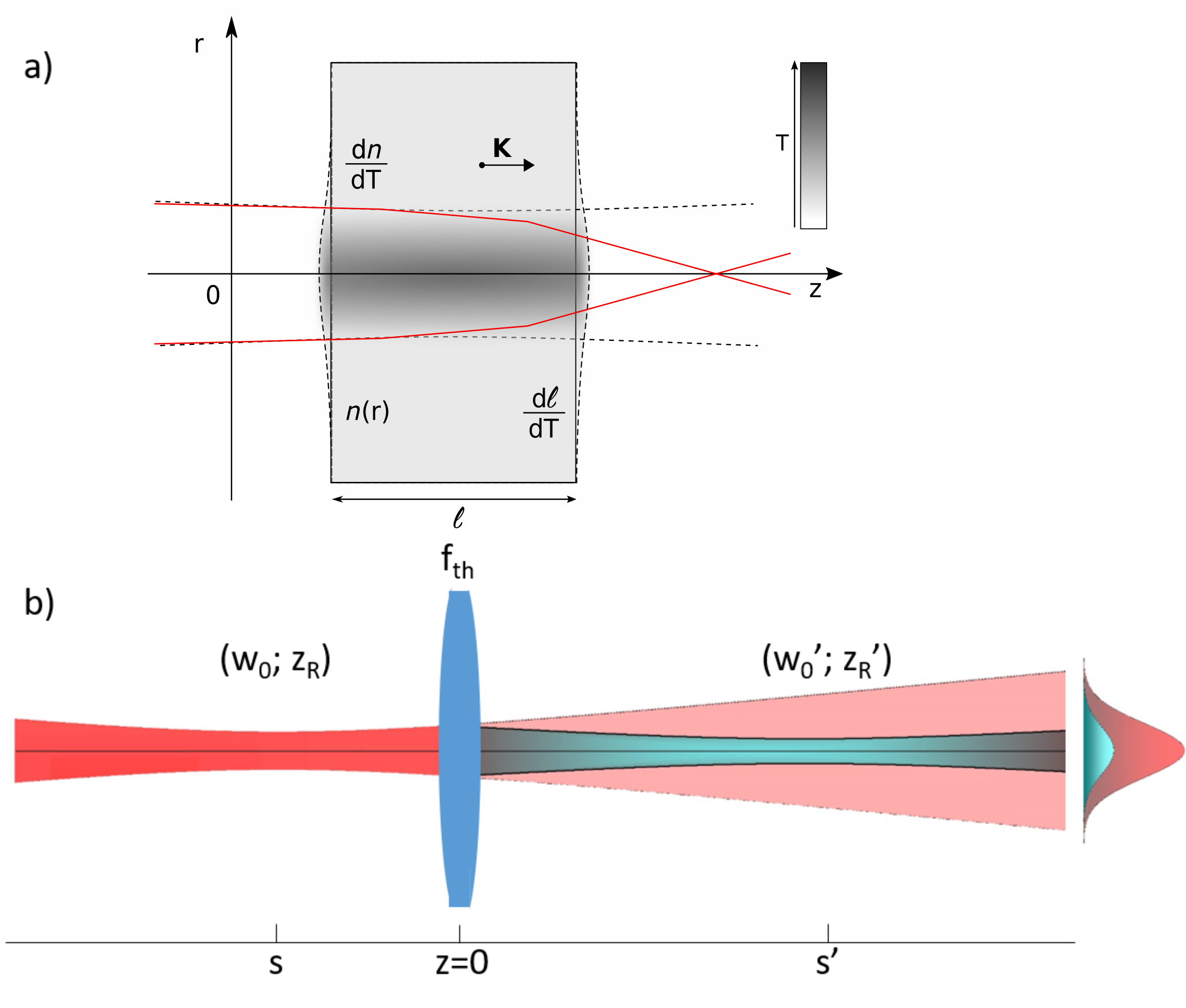}
\caption{\textbf{a)} Schematic visualization of thermal lensing of a Gaussian beam. Transmission of a laser beam through a partially absorbing medium of thickness $\ell$, and characterized by an absorption coefficient $b$, locally heats up the material at a rate set by its thermal conductivity $\kappa$. The Gaussian profile of the beam induces a temperature gradient that changes the refractive index, and hence the beam path, according to the temperature dependence $dn/dT$ of the substrate. Thermal expansion $d \ell/dT$ and strain dependence of the refractive index can further change the direction of wave propagation (\textbf{k}) in the medium, which acts as a thin, weak lens. \textbf{b)} Sketch of a thin lens $f_{th}$ positioned along the path of a Gaussian beam. The propagation of the incoming beam, characterized by a waist $w_0$ (and Rayleigh length $z_R$) placed at a distance $s$ from the lens, will be modified by $f_{th}$, that will create a new real (virtual) waist $w_0'$ at a distance $s'>0$ ($s'<0$) from the lens, according to Eq. \ref{ABCDa}. The sign convention for the object (image) position follows the one of ray optics: $s>0$ ($s'>0$) indicates a position on the left (right) of the lens plane.}
\label{Fig1}
\end{center}
\end{figure}

Since 1965, thermal lensing effects \cite{Gordon_1965} and more generally thermally induced wavefront distortions in high-power laser systems have been extensively investigated \cite{Sparks_1971, Bendow_1973, Sheldon_1982}. 
As already anticipated, such a phenomenon originates from the local heating caused by the transmission of a laser beam inside an optical element, which acts as a partially absorptive medium. Owing to the temperature dependence of the refractive index of the medium, the optical path experienced by the beam is modified in connection with the spatially inhomogeneous temperature distribution within the optical component, which acts as a "thermal lens" for the beam propagation, see Fig. \ref{Fig1}a.
Such a phenomenon encompasses a wide class of research fields and optical setups, spanning from high-energy laser physics to biological and material sciences. 
While thermal lensing may enable to devise various types of imaging techniques, such as the photo-thermal or thermal lens spectrometry employed for single-molecule detection of non-fluorescent compounds \cite{Uchiyama_2000,Tokeshi_2005}, it is generally an undesired effect in all cases where optimal beam profile quality of high-power lasers is sought \cite{Klein_1979, Klein_1996}.

Depending on the medium, thermal lensing can originate from different mechanisms, including thermal expansion of the material, strain and temperature dependence of the refractive index. This makes an effective description of the thermal lens associated to a generic system highly non-trivial. 
However, for optical materials such as quartz, fused silica or BK7 glass, and even more so for high purity optics with high damage thresholds, thermal lensing effects can be ascribed to the sole temperature dependence of the refractive index, $dn/dT$. 
In that case, neglecting contributions associated both with the volume expansion and mechanical stress of the material \cite{Waxler_1973}, and with the coating film deposited on the substrate \cite{Beausoleil_2003, Hello_1990}, thermal lensing of an optical element is quantified in terms of a thermal focal length $f_{th}$ that can be expressed as \cite{Bogan_2015}:

\begin{equation}
f_{th} = \frac{2 \pi \kappa}{1.3 b (dn/dT) \ell} \frac{w^2}{P} \equiv \frac{1}{m_0} \frac{w^2}{P}
\label{fth_formula}
\end{equation}
Here $P$ and $w$ denote the beam power and waist, respectively. $\kappa$ represents the thermal conductivity of the material,
$b$ its absorption coefficient, $dn/dT$ yields the temperature dependence of the refractive index, and $\ell$ the thickness of the medium. 
Namely, the optical element inducing thermal lensing can be considered as a thin lens whose focal length scales inversely with the incident intensity $I=2P/(\pi w^2)$, with a proportionality constant $m_0$ that depends on the specific properties of the substrate.
In particular, from Eq. \ref{fth_formula} one can see that for a given light intensity impinging on an optical element, $f_{th}$ will be larger, hence thermal lensing effects will be weaker, for those substrates that are thin, that feature low absorption and high thermal conductivity, with a refractive index weakly varying with temperature.

In order to gain an intuitive picture of thermal lensing effects within a generic setup, and to understand how they can be possibly cancelled out, it is useful to recall how a thin lens modifies the properties of an incident Gaussian beam \cite{Self_1983}.
Given an incoming beam featuring a waist (Rayleigh length) $w_0$ ($z_R= \pi w_0^2/ \lambda$) at a distance $s$ from a thin lens of focal length $f_{th}$, see sketch in Fig. \ref{Fig1}b, the focusing element will create a new waist $w_1$ at a distance $s'$, according to the following relations:

\begin{subequations}
\begin{align}
s' & = \frac{\frac{z_R^2}{f_{th}}-s(1-\frac{s}{f_{th}})}{\frac{z_R^2}{f_{th}^2}+(1-\frac{s}{f_{th}})^2} \\
\frac{w_1}{w_0} & = \frac{1}{\sqrt{(1-s/f_{th})^2+(z_R/f_{th})^2}}
\end{align}
\label{ABCDa}
\end{subequations}
From these relations, then, one can immediately notice the following facts:
First, for $f_{th}\rightarrow \pm \infty$ $s'=-s$ and $w_1=w_0$, i.e. the beam will not be modified.
Second, for any finite value of $f_{th}$, a new (real or virtual) beam waist will be created at a position that depends both upon the distance $s$ of the lens from the first waist, and on the initial beam parameters. 
As a consequence, the radius of curvature of the incoming beam, $R_0(z)=(z+s)(1+(\frac{z_R}{z+s})^2)$, will be modified according to $R(z)=(z-s')(1+(\frac{z_R'}{z-s'})^2)$ along the subsequent optical path. 
As a consequence, the far field intensity distribution of the beam will vary, enabling to quantify thermal lensing effects, for instance by measuring the change of the relative power transmitted through a slit placed behind the thermal lens, as a function of the incident power \cite{Sheldon_1982}. Alternatively, thermal lensing effects can be precisely characterized by coupling the beam to an optical cavity \cite{Bogan_2015}: The presence of thermal lenses along the beam path will be reflected into a sizable change in the coupling efficiency to the different cavity eigenmodes.  
These or similar techniques allow to retrieve the values of $f_{th}$ and $m_0$ associated with a given optical element, see Eq. \ref{fth_formula}, with no need to rely on a precise knowledge of the material properties.

Finally, in light of the forthcoming discussion in the next sections, it is useful to consider Eq. \ref{ABCDa} in the special case $s=0$, i.e. when the input beam waist lays on the plane of the thermal lens.
In this case, the position and size of the new waist become, respectively:
\begin{subequations}
\begin{align}
s' & = \frac{\frac{z_R^2}{f_{th}}}{1+\frac{z_R^2}{f_{th}^2}} \\
\frac{w_1}{w_0} & = \frac{1}{\sqrt{1+(\frac{z_R}{f_{th}})^2}}
\end{align}
\label{ABCD1a}
\end{subequations}
One can notice that, if $|f_{th}| \gg z_R$, by positioning the thermal lens in the beam focus the light propagation is modified only within a very small region behind the lens plane, while being unaffected at larger distances, since $s' \sim z_R^2/f_{th}$, and $w_1 \sim w_0$ up to corrections of the order of $(z_R/f_{th})^2$. Correspondingly, it is easy to check that the radius of curvature of the outgoing beam will coincide with the one of the incoming beam at all distances, aside for $O(z_R^2/|f_{th}|)$ corrections. 
Namely, whenever $|f_{th}| \gg z_R$, thermal lensing can be efficiently canceled by placing the substrate within a focus of the incoming beam \cite{Sheldon_1982, Uchiyama_2000, Silva_2011}. As it will be discussed more in detail in Section 4, this observation sets the basis for devising an optical trapping setup free from thermal effects.

\section{Characterization of thermal lensing within a model setup}

A prerequisite to minimize thermal lensing within a generic optical setup is to identify the main sources of such an effect by estimating the $f_{th}$ associated with each optical element traversed by the laser beam.
Since any material unavoidably introduces some degree of thermal phase aberrations, when designing a high-power optical setup it is in general desirable to minimize the number of components the laser beam has to pass through.
For this reason, our optical dipole trap design employs as few optical elements as possible to adjust the beam power and waist on the atoms: Neglecting all reflective elements, our design (see sketch in Fig. \ref{Fig2}a) is solely composed by three lenses, one AOM and the quartz window of the vacuum chamber, within which the atomic clouds are produced.

The ODT light source is provided by a $YLR-300$ multimode fiber laser module by IPG Photonics delivering up to $300\, \mathrm{W}$ output power. The central emission wavelength is $1070 \, \mathrm{nm}$ and the output beam waist is $w_0\simeq 2.2(2) \,\mathrm{mm}$ with negligible ellipticity. Due to the clear aperture of the AOM of about $2.5 \times 1.75\, \mathrm{mm^2}$, two lenses are employed to de-magnify the beam waist down to about $550\, \mathrm{\mu m}$. 
The first order diffracted beam of the AOM is then re-expanded in order to obtain a waist $w_3\simeq2200 \,\mathrm{\mu m}$ on the last lens $f_3=250 \,\mathrm{mm}$, employed to focus the beam down to a waist of about $w_{at}\simeq 45 \,\mathrm{\mu m}$ on the atomic cloud after passing through the vacuum chamber window.
All lenses employed in our design are one inch UV fused silica elements with anti-reflection V-coating at $1064$/$532\, \mathrm{nm}$\footnote{ UVFS YAG-ML lens by $Thorlabs$}. These represent a cheap, convenient option for high power applications, due to a very small $dn/dT\simeq12\cdot10^{-6} \, \mathrm{K^{-1}}$ \cite{Toyoda_1983} and a low thermal expansion coefficient $\alpha \simeq 0.5\cdot 10^{-6}\, \mathrm{K^{-1}}$ \cite{Hahn_1972}. The $CF-40$ window of the vacuum chamber is instead made by a $3.3\, \mathrm{mm}$ thick quartz substrate with custom anti-reflection coating \footnote{ AR-coating $426 \,\mathrm{nm} + 532 \,\mathrm{nm} + 630-675 \,\mathrm{nm} + 1064\,\mathrm{nm}/0^\circ$ by \emph{LaserOptik Garbsden}}. 
Finally, the AOM is realized by a 31 $\mathrm{mm}$ thick $TeO_2$ crystal \footnote{ $3110-191$ by \emph{Gooch\&Housego}}.

\subsection{Methods}

\begin{figure}[t!]
\begin{center}
\includegraphics[width=8cm]{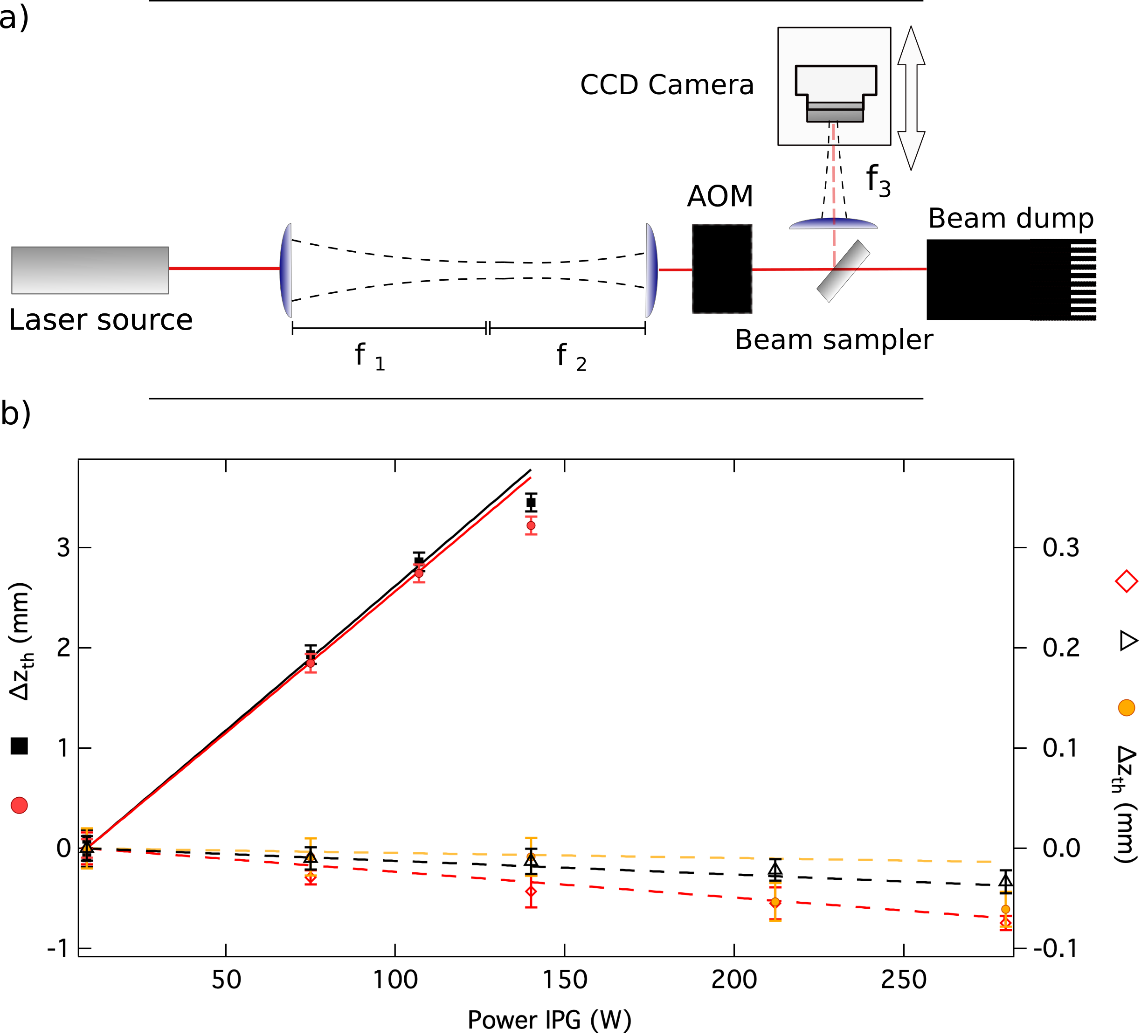}
\caption{Characterizing thermal lensing of different optical elements. \textbf{a)} Setup for thermal lensing measurements. 
Along the full path, the high power beam passes through two lenses and one AOM. A $BSF10-C$ coated beam sampler enables to create a low-power ($P<10\, \mathrm{W}$) copy of the beam, which is focused by a third lens $f_{3}$ and sent to a CCD camera mounted on a translation stage (double arrow). The focus position is measured by recording the peak intensity of the Gaussian spot versus the camera position.
\textbf{b)} Thermal shifts $\Delta z_{th}$ as a function of the laser power recorded for different combinations of optical elements.
Right axis: $\Delta z_{th}$ due to the $f_1-f_2$ telescope with $f_2=50\, \mathrm{mm}$ in $Suprasil$\textsuperscript\textregistered $3001$ (black triangles) or in UV fused silica (red diamonds). 
The shift of the $f_1=200\, \mathrm{mm}$ fused silica lens alone (yellow circles) has been tested directly by measuring its focus shift versus the beam power. 
For each data set, the dashed line is the corresponding shift calculated by Gaussian beam propagation analysis, assuming each element to represent an additional lens with$f_{th}$ given by Eq. \ref{fth_formula} and characterized by the corresponding $m_0$ value listed in Table \ref{table1}. 
Left axis: Thermal shift of the optical setup with inclusion of the AOM crystal, with (black squares) or without (red circles) quartz window in the beam path.  
The AOM was placed at $d_{AOM,2}= 3(1) \, \mathrm{cm}$ behind the second lens $f_2$, the last lens $f_3$ at $d_{3,AOM} = 58(2) \, \mathrm{cm}$, whereas the window (if present) was at $d_{win,3}= 12(1)\, \mathrm{cm}$ after $f_3$. 
Solid lines (same color code) show the focus shift calculated by Gaussian beam propagation analysis, assuming the AOM thermal lens to be described by Eq. \ref{fth_formula} with the $m_0$ value given in Table \ref{table1}.}  
\label{Fig2}
\end{center}
\end{figure}

As already discussed in the previous section, one method to measure the thermal lens of an optical element is to monitor the beam divergence behind it. This can be done by inspecting how the axial intensity profile of the outgoing beam
\begin{equation}
I(z, z_0)=I_0 (\frac{w_0}{w(z)})^2=\frac{I_0 }{1 + (\frac{z-z_0}{z_R})^2}
\label{fit}
\end{equation}
depends upon the power impinging on the substrate.
Here $I_0$, $w_0$ and $z_0$ denote the maximum intensity, the waist size and position, respectively, all affected by the specific thermal lens of the examined optical element. 
For the case of one single lens along the optical path, the axial intensity profile for a given power level can be measured by focusing the beam on a CCD camera, moved along the propagation axis through a translation stage. For each $z$ position, the intensity $I(z)$ can be then obtained as the amplitude of the laser spot, extracted from a two-dimensional Gaussian fit.
For the case of several elements, the generalized scheme depicted in Fig. \ref{Fig2}a can be employed.

In spite of the conceptual simplicity of this method, we emphasize that special care must be taken to avoid systematic effects connected with the need to attenuate the beam intensity on the detector. While beam powers exceeding 100 W are needed to reveal sizable thermal aberrations induced by the optical elements composing our setup, already a few milliwatts saturate the CCD camera chip. This implies the need of a filtering stage, whose associated thermal lens can easily invalidate the whole measurement.
To this end, we found that a filtering stage that limits additional strong thermal aberrations can be realized by first sending the high power beam to a $BSF10-C$ coated beam sampler, from which a beam with power lower than 10 W is derived. 
After this point thermal effects are negligible, and a second attenuation stage can be safely obtained by letting the beam cross a high-reflection mirror before hitting the CCD sensor. Yet, the thermal lens of such attenuation stage remains significant when considered in combination with optical elements featuring very long $f_{th}$.

By employing such a simple setup we recorded, for various laser power levels and different combinations of optical elements, the corresponding axial intensity profiles which, fitted to the trend given by Eq. \ref{fit}, provided the focus position, with an uncertainty essentially dominated by the intensity fluctuations of the spot on the CCD camera. 
Thermal lensing of the elements placed within the beam path was then quantified in terms of the shift $\Delta z_{th}$ of the focus location $z_0$, relative to the one recorded under low power conditions. 
We underline that, owing to the minimum time resolution of our CCD camera, $\delta t = 50\,\mathrm{ms}$, we did not attempt a dynamical characterization of thermal lensing, and all the data reported in the following have been recorded in stationary conditions. 

\subsection{Results}
First of all, we looked at the thermal lens associated with one single fused silica lens $f_1=200 \mathrm{mm}$ placed in front of the laser output, at a distance $L_1 = 60(5)\,\mathrm{mm}$, 
 much smaller than the Rayleigh length associated with the output waist $w_1 \simeq 2.2\,\mathrm{mm}$.
By following the scheme previously described, we measured the shift $\Delta z_{th}$ of the focus position, relative to the location of a low power ($P<10\,\mathrm{W}$) beam. The resulting trend, recorded as a function of the incident power, is presented in Fig. \ref{Fig2}b as yellow circles. 
As one can notice, thermal effects associated with $f_1$ together with the filtering stage cause only very small shifts of the focus position, $\Delta z_{th}$ remaining below $80\,\mathrm{\mu m}$ up to the highest power of $280 \,\mathrm{W}$ (intensity of about $3.7\,kW/cm^2$).

By following a similar procedure, we quantified the thermal lens generated by two lenses $f_1=200\, \mathrm{mm}$ and $f_2=50 \, \mathrm{mm}$ in a de-magnifying $1:4$ telescope configuration. The thermal effects of the telescope were monitored by measuring the shift of the focus produced by a third lens $f_3$, positioned within the low power region behind the beam sampler, hence yielding a negligible contribution to thermal aberrations, see sketch in Fig. \ref{Fig2}a.
Due to the 4-fold de-magnification of the beam, $w_2=w_1/4\simeq 550\,\mathrm{\mu m}$, the second lens experienced a 16-fold increased intensity, relative to the one impinging on $f_1$. 
The resulting trend of $\Delta z_{th}$ is shown in Fig. \ref{Fig2}b, for a second lens $f_2$ made of fused silica
\footnote{ LA4148-YAG-ML by $Thorlabs$} 
(red diamonds) or $Suprasil$\textsuperscript\textregistered $3001$ \footnote{AR/AR1070 PLCX-25.4/25.8 S3001 by $Laser Components$} 
(black triangles), respectively.
In spite of the sizable increase of the intensity on the second lens of the telescope, in both cases thermal lensing causes only negligible shifts of the $f_3$ focus location, $\Delta z_{th}\lesssim 100 \,\mathrm{\mu m}$. 
Given that the atom clouds initially loaded within the ODT feature sizes easily exceeding a few millimeters, all these variations are irrelevant for our purpose, and a quantitative analysis of these three data sets goes beyond the scope of the present work. 
Nonetheless, in relation with the $f_1-f_2$ data, we remark how our simple method indeed enables to distinguish among the (weak) thermal lenses of the two different substrates, the $Suprasil$\textsuperscript\textregistered lens clearly outperforming the fused silica one.
Further, we stress that the single lens data set cannot be directly compared with the one taken with the telescope owing to the different setup.
In particular, as it will be discussed in the following, the former characterization was affected by stronger spurious effects associated with thermal lensing due to the filtering stage.

\begin{table}[h!]
\centering
 \begin{tabular}{||p{3.8cm} | p{3.4cm} | p{1.4cm}||} 
 \hline
 Optical element/substrate & $m_0 (\mathrm{m W^{-1} }$) & Reference\\ [0.5ex] 
 \hline\hline
  AOM/ $TeO_2$ crystal & $-1.13(7) \times 10^{-10}$ & \cite{Bogan_2015}\\
 \hline
 Window/ Quartz & $-4.9(5) \times 10^{-13}$ (o-axis)
 
  $-10.1(11) \times 10^{-13}$ (e-axis) & \cite{Toyoda_1983}\\ [1ex]
  \hline
  Lenses/ UV fused silica & $4.1(8) \times 10^{-12}$ & \cite{Khashan_2001} \\ \hline
  Lenses/ $Suprasil$\textsuperscript\textregistered & $10(1) \times 10^{-14}$ & \cite{Leviton_2015} \\
 \hline
\end{tabular}
\caption{List of $m_0$ values characterizing the different sources of thermal lensing in our setup. The specified uncertainties combine the ones given in the corresponding references with the uncertainty in the determination of the specific substrate thicknesses.}
\label{table1}
\end{table}

As a next step, we characterized the thermal lens associated with the acousto-optic modulator which enables to control the beam power on the atomic sample. 
In particular, we considered a standard AOM \footnote{$3110-191$ by \emph{Gooch\&Housego}} made by an $AR$-coated $TeO_2$ crystal that enables maximum diffraction efficiencies around $85\%$ for an input beam waist of $550\, \mathrm{\mu m}$. 
To this end, we positioned the AOM a few cm after the $f_1-f_2$ telescope, see sketch in Fig. \ref{Fig2}a, and applied the same protocol discussed above for the telescope characterization. The outcome of this study is presented in Fig. \ref{Fig2}b as black squares.
Despite our working conditions were far from the AOM damage threshold of $10\, \mathrm{MW/cm^2}$ at $1070\, \mathrm{nm}$, the $TeO_{2}$ crystal resulted to provide a shift of the focus location about two orders of magnitude larger than the ones observed with the lenses alone.
Given that the observed focal shift appear to be only weakly modified by the presence of an additional quartz window behind the AOM, see red circles in Fig. \ref{Fig2}b, we conclude that the only sizable source of thermal lensing in such a model setup is represented by the $TeO_2$ crystal. 
Additionally, it is interesting to notice how the shift caused by the AOM is opposite to the one observed with the other elements, signaling a negative $dn/dT$ of the $TeO_2$ substrate.

Our findings, despite not enabling an accurate, independent measure of the $m_0$ parameters characterizing all elements of the setup, appear compatible with the values that can be found in literature \cite{Bogan_2015,Toyoda_1983,Khashan_2001,Leviton_2015} for the different substrates. 
This was verified by comparing the experimental data with the outcome of simulations of Gaussian beam propagation, shown as dashed and solid lines in Fig. \ref{Fig2}b. Our analysis assumed each thermal lens to be describable as a thin lens positioned in correspondence of the associated physical substrate, and characterized by the $m_0$ values retrieved from previous studies, summarized in Table \ref{table1}. 
In particular, the simulated $\Delta z_{th}$ quantitatively match all experimental data sets, except for the case of one single fused silica lens, for which the measured shift (yellow circles) significantly exceeds the simulated one (yellow dashed line).
We ascribe such a sizable mismatch, absent when considering two fused silica lenses in a telescope configuration (red diamonds), to the spurious contribution of the thermal lens associated with the filtering stage that, for the single lens measure, was illuminated by a tightly focused beam.

Consistently with the trends presented in Fig. \ref{Fig2}b, one can notice from Table \ref{table1} how the focal length $f_{th}$ associated with the $TeO_2$ substrate is negative and about $25$ ($200$) times shorter than the one of fused silica (quartz) elements under the same intensity conditions. This confirms that the AOM crystal represents the major and only relevant source of thermal lensing within our ODT setup. 
Based on the results of Ref. \cite{Bogan_2015} and on our measurements, the AOM is expected to feature $|f_{th}| \leq 10\, \mathrm{m}$ for the maximum power delivered by our source and with a $550\, \mu\mathrm{m}$ beam waist, whereas all other elements exhibit ten or hundred times longer thermal focal lengths. 
From a simple Gaussian beam propagation analysis, it is easy to verify that the $f_{th}$ of a $TeO_2$ crystal, when placed behind a de-magnifying telescope as in typical optical trapping setups, may cause a few millimeters thermal shift of the focus of the last lens $f_3$. 
On the other hand, we remark that the contribution of other elements, irrelevant within the setup under consideration in the present study, could become important when illuminated with much higher intensities.  
We finally emphasize that special care must be taken in the alignment of the beam at the center of the AOM crystal and the other optical elements. This is essential to guarantee paraxial working conditions and to avoid, besides thermal shifts of the waist position, subject of the present study, thermal induced aberrations that easily lead to strong astigmatism, especially when few micron beam waists are considered.

\section{Compensation of thermal lensing effects}

\begin{figure}[t!]
\begin{center}
\includegraphics[width=8cm]{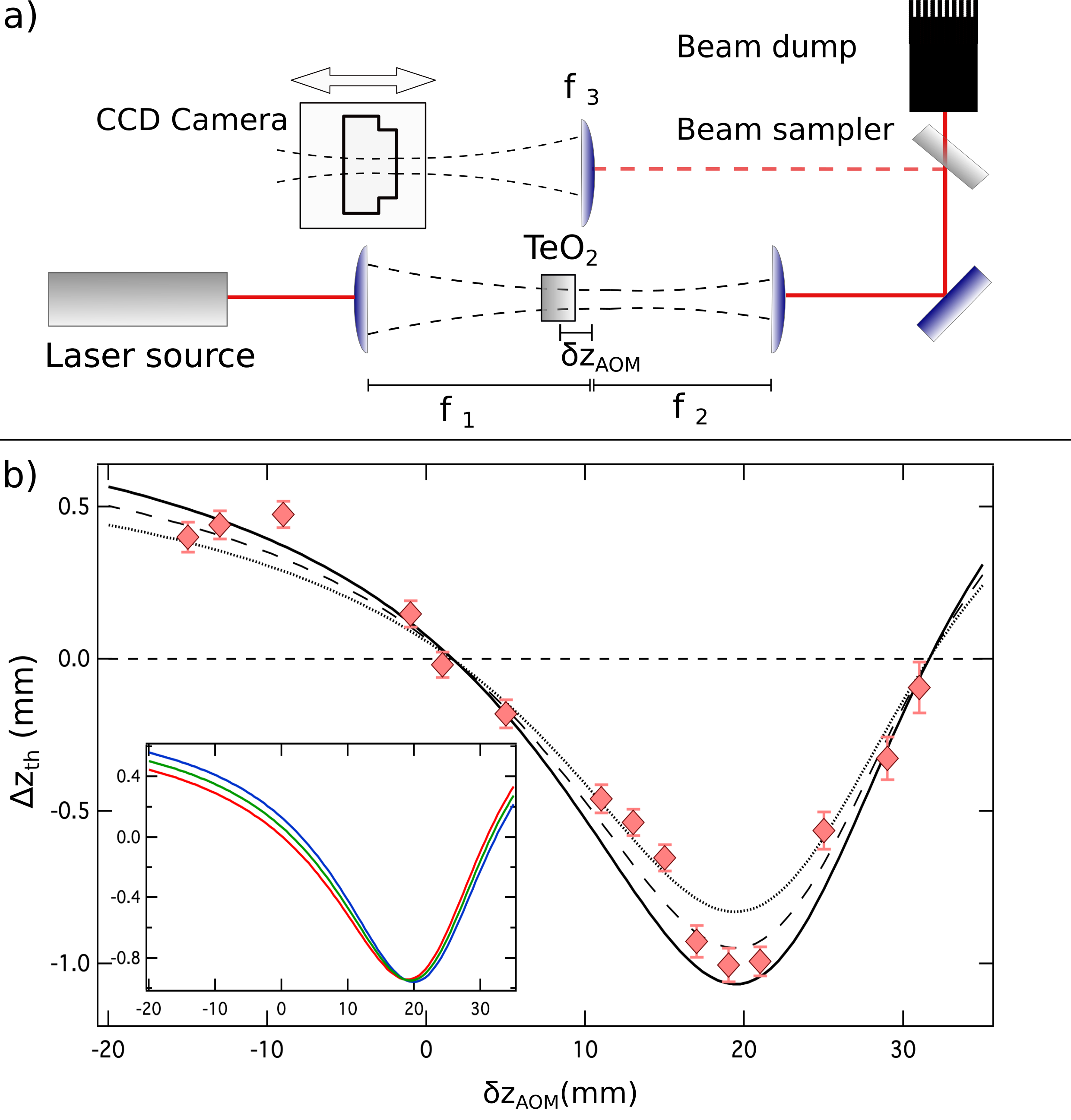}
\caption{Model setup to control thermal lensing effects.
\textbf{a)} Schematic view of the optical scheme employed for the characterization of the AOM thermal lens, as a function of the crystal position. A $TeO_2$ crystal is placed at a variable distance $\delta z_{AOM}$ from the focus within the $f_1-f_2$ telescope as shown in the picture. For this measure, $f_1=300\, \mathrm{mm}$ and $f_2=75\, \mathrm{mm}$. The $f_3$ lens is placed at $d_{3,2}=47(2)\, \mathrm{cm}$ from the second lens $f_2$, and the focus location is monitored for different levels of incident laser power through a CCD camera. 
\textbf{b)} Thermal-induced shift $\Delta z_{th}$ of the $f_3$ focus position experimentally determined (red diamonds), as a function of the AOM distance from the $f_1$ focus. $\Delta z_{th}$ is obtained by comparing high and low power data acquired at $P=50(1)\, \mathrm{W}$ and $P=9.0(5)\, \mathrm{W}$, respectively. 
The shift predicted by the Gaussian beam propagation analysis is shown as black lines for $P=55\, \mathrm{W}$ (solid), $P=50\, \mathrm{W}$ (dashed) and $P=45\, \mathrm{W}$ (dotted). 
Inset: expected behavior of $\Delta z_{th}$ for an incident power of $55\, \mathrm{W}$ for three different distances between second and third lens: $d_{3,2}=47\, \mathrm{cm}$ (green), $d_{3,2}=50\, \mathrm{cm}$ (blue) and $d_{3,2}=44\, \mathrm{cm}$ (red).}
\label{Fig3}
\end{center}
\end{figure}

As anticipated when discussing Eq. \ref{ABCDa} and the special case described in Eq. \ref{ABCD1a}, the impact of one thermal lens on a propagating beam can be minimized by positioning the thermal element within a focus along the optical path \cite{Sheldon_1982, Uchiyama_2000}.
In particular, this is possible whenever the thermal focal length greatly exceeds the Rayleigh length of the incoming beam, $|f_{th}| \gg z_R$, which is actually fulfilled by the typical trapping setups in cold atom experiments. 
Indeed, the $f_{th}$ connected with the $TeO_2$ crystal of the setup is such that $|f_{th}|/z_R > 10$ for the highest intensities explored in this study. As a first step in the direction of eliminating the effect of the AOM thermal lens on the trapping beam, we characterized how the focus produced by $f_3$ on a CCD camera, see sketch in Fig. \ref{Fig3}a, shifts as a function of the position $\delta z_{AOM}$ of a $TeO_2$ crystal relative to the focus of the $f_1-f_2$ telescope, for two different values of the incident power (see details in Fig. \ref{Fig3} caption). Given that $f_1$ focuses the incident beam down to waists of about $45 \, \mathrm{\mu m}$, the power level was in this case kept below $60 \, \mathrm{W}$. Nonetheless, this corresponds to an intensity on the AOM crystal about 40 times higher than the one reached within standard operating conditions, yielding $f_{th} \sim 30 \, \mathrm{cm}$.  
The acquired data are shown as red diamonds in Fig. \ref{Fig3}b, together with the simulated curves obtained from the analysis of Gaussian beam propagation. The simulation accounted for the three lenses of the setup, placed at fixed positions, and it included a thin lens $f_{th}$ at the center of the AOM crystal, characterized by the $m_0$ parameter reported in Table \ref{table1}.
From Fig. \ref{Fig3}b one can notice how a small variation of the $TeO_{2}$ thermal lens position, by less than the crystal thickness, may strongly modify the beam propagation, leading to both positive and negative shifts of the $f_3$ focus with the incident power on the $TeO_{2}$ crystal. 
Notably, the overall trend of $\Delta z_{th}$ is reproduced by our simple theoretical analysis, implying that, for our typical working conditions, Eq. \ref{fth_formula} provides an excellent approximation to describe the thermal lenses of our setup.
As shown in the inset of Fig. \ref{Fig3}b, the overall trend of $\Delta z_{th}$ exhibits a much weaker dependence upon the distance $d_{3,2}$ between the second and the third lens. This can be understood considering that the beam behind the telescope has a Rayleigh length of the order of one meter, much larger than the one featured by the beam within the focus of the telescope, on the order of $3 \, \mathrm{m m}$.

Based on the experimental data and the simulation results shown in Fig. \ref{Fig3}b, one can notice that thermal lensing can be zeroed for two, rather than one, distinct AOM positions. Indeed, besides the $\delta z_{AOM}=0$ configuration, negligible thermal shifts were also observed for $\delta z_{AOM} \sim 30 \, \mathrm{mm}$. 
By inspecting the simulated beam propagation through the whole setup sketched in Fig. \ref{Fig3}a, we found that this second $\Delta z_{th}=0$ point occurs for a position of the AOM that yields, at the plane of the third lens, a radius of curvature that coincides with the one of the unperturbed beam, obtained for $|f_{th}|=\infty$.
While also this second configuration enables to strongly suppress thermal lensing, it is however less robust than the $\delta z_{AOM}=0$ one.
Given that in this case the radii of curvature associated with different power levels coincide only at the $f_3$ plane, rather than throughout the whole optical path, the beam magnification due to $f_{th}$ at the $f_3$ plane may significantly differ from unity. As a consequence, although the position of the focus produced by $f_3$ will only weakly depend upon the specific value of $f_{th}$ (i.e. of incident power), the beam waist may sizable vary, relative to the $|f_{th}|= \infty$ case. 
 
Aside for understanding the detailed behavior of $\Delta z_{th}$, this proof-of-principle experiment shows that it is indeed possible to cancel out the thermal lensing effect introduced by the AOM by properly adjusting its position to match a beam waist along the optical path.
Importantly, this holds irrespective of the systematic uncertainty in the determination of $\delta z_{AOM}$ within the optical setup and, possibly, of a small deviation from the perfect $f_1-f_2$ telescope configuration.
On the other hand, the present configuration cannot be employed in a realistic optical trapping setup. Indeed, the beam waist in the focus of the $f_1-f_2$ telescope is about $45\, \mathrm{\mu m}$, which drastically reduces the diffraction efficiency of the $TeO_2$ crystal, and that would yield at the highest power level of our laser source an intensity exceeding the AOM damage threshold. 

In order to overcome this issue while keeping the $TeO_2$ crystal within a focus of the optical trapping beam, among different solutions, we opted for a scheme based on the same elements depicted in Fig. \ref{Fig3}a, with the first two lenses no longer in a telescope configuration but rather acting as an equivalent lens with effective focal length $f^{eq}_{1,2}$. The latter will generally depend upon the parameter $\delta z$, defined as:
\begin{equation}
\delta z= L_2-L_1 -(f_1+f_2)
\label{collimated condition}
\end{equation}
Here $L_i$ and $f_i$ denote the position and the focal length of the $i-th$ lens, respectively.
The first lens was mounted on a translation stage with a resolution of $10^{-2}\, \mathrm{mm}$, and the position of the focus produced by $f^{eq}_{1,2}$ was determined by Gaussian beam matrices as a function of the $L_1$ position, hence of $\delta z$.

From our theoretical analysis we found that there exist various $L_1$ configurations, all for small and positive $\delta z$ values, yielding a focus at relatively short distances from the second lens $f_2$, with the beam waist ranging between $550$ and $500 \, \mu \mathrm{m}$.
Therefore, we proved the feasibility of such a scheme by fixing the AOM crystal at two different representative distances $d_{AOM,2}$ from the second lens $f_2$: $d_{AOM,2}=23(2)\, \mathrm{cm}$ and $d_{AOM,2}=3(1)\, \mathrm{cm}$, respectively. In particular, the latter one corresponds to the focus position of the equivalent lens with $\delta z \simeq 0$, i.e. with the two lenses $f_1-f_2$ very close to the collimated condition.
At this point, and for each of the two AOM configurations, we finely scanned $\delta z$ upon varying the position $L_1$ of the first lens, hence modifying the resulting $f^{eq}_{1,2}$ and the associated focus location. 
This procedure is less intuitive than the one previously described when discussing Fig. \ref{Fig3}b data, since the change in position of the first lens, rather than the AOM one, simultaneously modifies the focal length $f^{eq}_{1,2}$ and the position of the focal point relative to the $TeO_2$ crystal. 
On the other hand, this method has the advantage that it does not affect the alignment of the optical path behind the AOM once the diffracted first order beam is employed, as in standard working conditions of the trapping setup.
Despite this slightly modified measuring protocol, thermal effects arising from the AOM crystal could be quantified by monitoring how the focus produced by the third lens $f_3$ varied with $\delta z$ for two different levels of incident power, similarly to what discussed above the data shown in Fig. \ref{Fig3}. 

\begin{figure}[t!]
\begin{center}
\includegraphics[width=7cm]{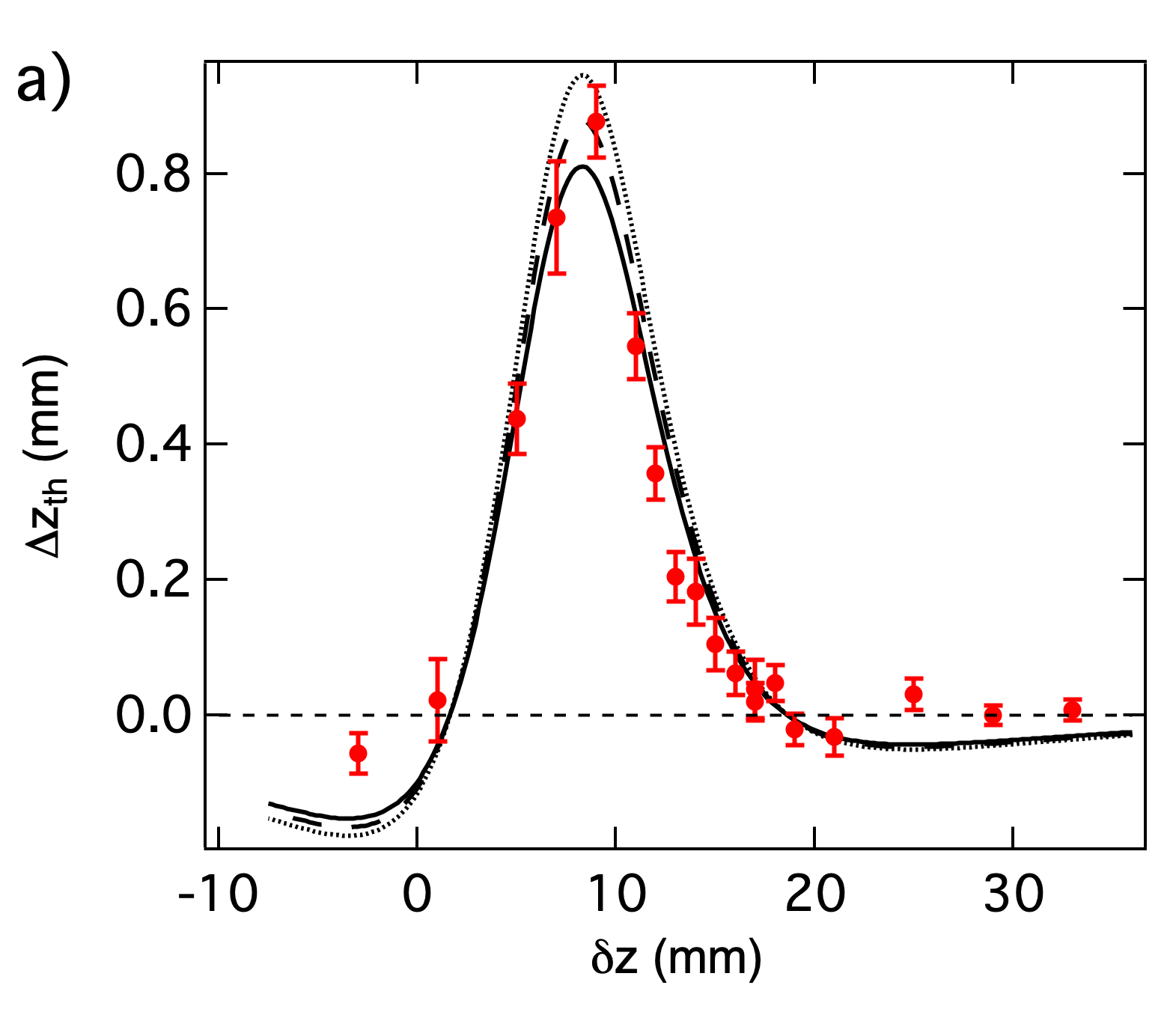}
\includegraphics[width=7cm]{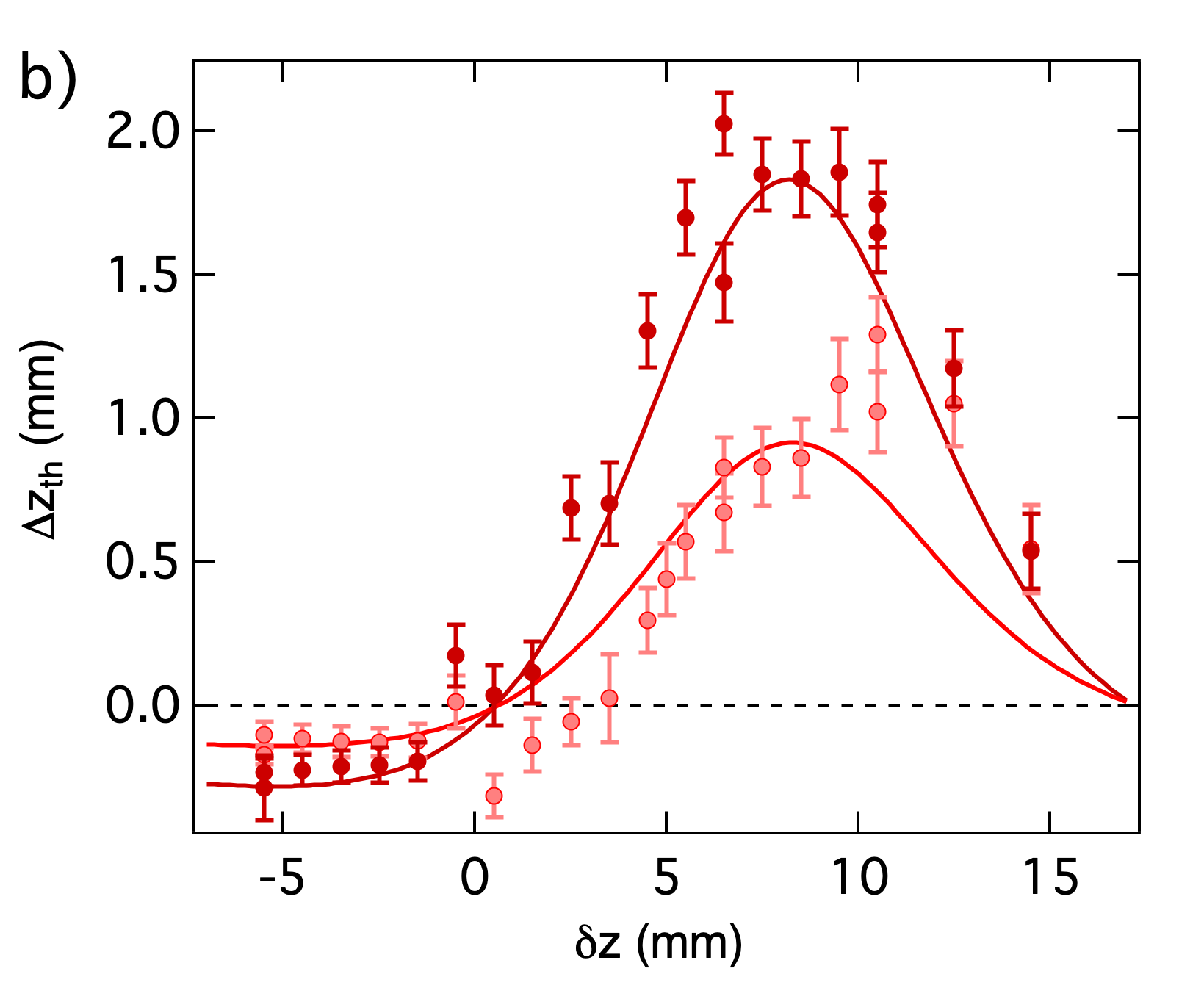}
\caption{Controlling the AOM thermal lensing through an equivalent lens. The two panels show the measured thermal shift $\Delta z_{th}$ (red circles) of the focus created by the last lens $f_3$ as a function of the parameter $\delta z$ given by Eq. \ref{collimated condition}, for the two different AOM locations discussed in the main text.
\textbf{a)} The AOM was positioned at $d_{AOM,2}=23(2)\, \mathrm{cm}$ relative to the plane of the second lens.
$\Delta z_{th}$ was obtained by comparing the focus position measured at $P=75(1) W$ and $P=9.0(5) W$, respectively.  
Black lines show the simulated $\Delta z_{th}$ for different high power levels: $80\, \mathrm{W}$ (dotted lines), $75\, \mathrm{W}$ (dashed lines) and $70\, \mathrm{W}$ (solid lines). 
\textbf{b)} Experimentally measured thermal shift as in panel a), but with the AOM positioned at $d_{AOM,2}=3(1)\, \mathrm{cm}$.  
Two high power values have been checked, relative to the low power reference at $P=9.0(5)\, \mathrm{W}$: $80(1)\, \mathrm{W}$ (light red circles) and $150(2)\, \mathrm{W}$ (dark red circles). Solid lines show the simulated trend expected for the two power levels.
For both data sets, $f_1=300\, \mathrm{mm}$ and $f_2=75\, \mathrm{mm}$, and the last lens $f_3$ was kept fixed at $d_{3,2}=155(2)\, \mathrm{cm}$.
In both panels, error bars combine the standard error of the axial intensity profile fitted to Eq. \ref{fit} for the high and low power data sets.  
}
\label{Fig4}
\end{center}
\end{figure}

The results of this latter characterization are presented in Fig. \ref{Fig4} for the two $d_{AOM,2}$ values considered here. 
In particular, Fig. \ref{Fig4}a shows the thermal shifts measured with the AOM positioned at $d_{AOM,2}=23(2)\, \mathrm{cm}$ from the second lens, whereas Fig. \ref{Fig4}b presents the outcome of the analogous characterization for $d_{AOM,2}=3(1)\, \mathrm{cm}$. 
For both AOM positions explored, the last $f_3$ lens was kept at a fixed distance $d_{3,2}=155(2)\, \mathrm{cm}$ from the second one.
Aside for slight quantitative changes, the observed trends of $\Delta z_{th}$ qualitatively agree with the one obtained when moving the $TeO_2$ crystal within the focus of the $f_1-f_2$ telescope, see Fig. \ref{Fig3}b. 
Also in these cases, the measured thermal shifts appear to be reasonably reproduced by our theoretical analysis, featuring a sharp peak connected via two zero-crossing points to two outer regions characterized by a slowly-varying value of $\Delta z_{th}<0$.
In both cases the range of $\delta z$ that can be investigated experimentally is limited on one side by the diffraction efficiency (too small beam waists on the AOM) and the finite $TeO_2$ crystal size on the other. 
These data demonstrate that even in this case it is possible to experimentally identify special configurations of the $f_1-f_2$ setup for which the thermal lensing effect of the $TeO_2$ crystal can be zeroed, while guaranteeing an AOM diffraction efficiency exceeding 80$\%$.

\begin{figure}[t!]
\begin{center}
\includegraphics[width=9cm]{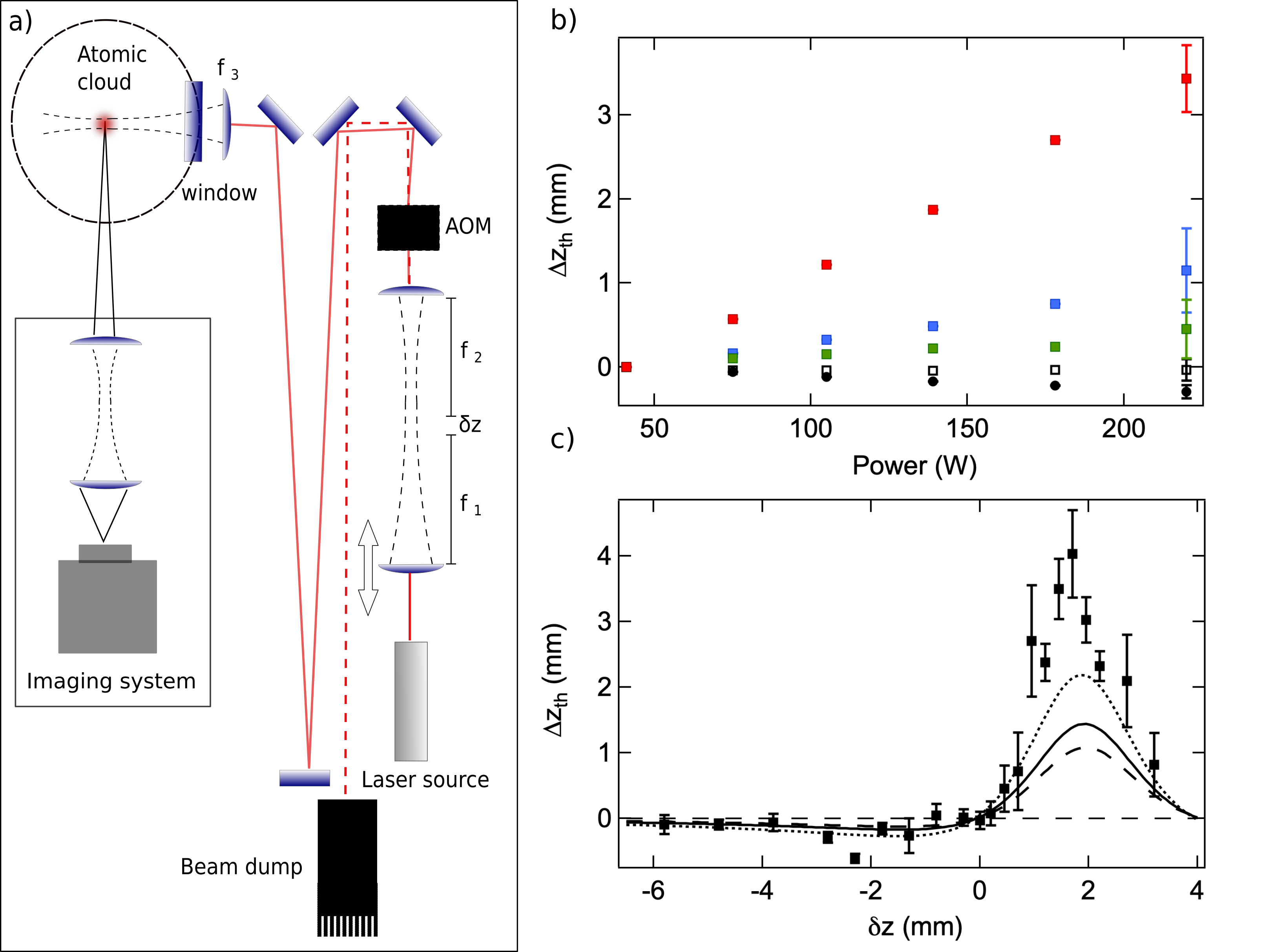}
\caption{Realizing an optical dipole trap without thermal lensing effects. 
\textbf{a)} Sketch of the setup employed to investigate thermal lensing by monitoring an atomic cloud trapped in the ODT. Typical atom number in the ODT after $400\, \mathrm{ms}$ illumination time ranges from $1\times10^5$ ($P=40\, \mathrm{W}$) to $7\times10^5$ ($P=220\, \mathrm{W}$). 
For this measurements, $f_1=200\, \mathrm{mm}$ and $f_2=50\, \mathrm{mm}$ while the last lens $f_3$ is placed at $d_{3,AOM}=200(5)\, \mathrm{cm}$.
\textbf{b)} Thermal shift $\Delta z_{th}$ of the focus position as a function of the beam power $P$ for different values of the parameter $\delta z$: $\delta z=1.7(1)\, \mathrm{mm}$ (red squares), $\delta z=1.2(1)\, \mathrm{mm}$ (blue squares), $\delta z= 0.45(1)\, \mathrm{mm}$ (green squares), $\delta z=0.0(1)\, \mathrm{mm}$ (white squares) and $\delta z=-2.3(1)\, \mathrm{mm}$ (black circles).
\textbf{c)} Thermal shift $\Delta z_{th}$ as a function of the parameter $\delta z$ at a fixed power $P=220(2)\, \mathrm{W}$. The solid line shows the thermal shift expected from the Gaussian beam matrices calculation considering the $f_{th}$ of the AOM crystal given by Eq. \ref{fth_formula} with the $m_0$ value shown in Table \ref{table1}. The dashed (dotted) line shows the expected thermal shift for $f_{th}+\Delta f_{th}$ ($f_{th}-\Delta f_{th}$) where $\Delta f_{th}$ is our estimate of $f_{th}$'s uncertainty of around $35\%$.
Error bars combine the statistical uncertainties of the high and low power reference data sets on the atomic cloud barycenter, obtained for each point from an average of $4$ independent measurements.}
\label{Fig5}
\end{center}
\end{figure}

We finally tested the efficacy of our scheme by directly monitoring the axial position of a cold atomic cloud confined within the high power beam, employing a configuration of the optical setup analogous to the one considered in Fig. \ref{Fig4}b, with $d_{AOM,2}=3(1)\, \mathrm{cm}$, see sketch in Fig. \ref{Fig5}a. 
By following procedures that will be described elsewhere \cite{Neri_2019}, we produced cold clouds of about $2.0(2)\times 10^8$ $^6Li$ atoms at $T\simeq80\, \mathrm{\mu K}$, which we subsequently illuminated with the ODT beam.
After an illumination time of $400\, \mathrm{ms}$, long enough to ensure that stationary conditions were attained, the position of the trapped sample along the ODT axis was obtained by Gaussian fits to the atomic density profiles, obtained through \textit{in situ} absorption imaging performed along one direction perpendicular to the trapping beam, see Fig. \ref{Fig5}a. 
In turn, for any value of incident power and of $\delta z$, the axial barycenter of the atom cloud reflects the waist position of the ODT beam, corresponding to the energy minimum of the optical potential. Fig. \ref{Fig5}b shows examples of the experimentally determined shifts of the cloud position along the beam axis, relative to the one obtained at the lowest possible power enabling to capture a detectable atomic fraction ($P=40(1)\, \mathrm{W}$), as a function of the power level for different $\delta z$ values. Also, these data show that one can adjust the $\delta z$ parameter to induce either positive or negative thermal shifts of variable magnitude and, most importantly, to cancel them out.

Finally, Fig. \ref{Fig5}c shows, as a function of $\delta z$, the thermal shift obtained by comparing the atomic cloud positions recorded under high ($P=220(2)\, \mathrm{W}$) and low ($P=40(1)\, \mathrm{W}$) power conditions. The resulting trend qualitatively matches the one presented in Fig. \ref{Fig4}, albeit featuring a poorer agreement with the simulation (solid line). In particular, our theoretical model systematically underestimates the measured shifts (black squares) around the region of maximum $\Delta z_{th}$, even when allowing for a $\pm 35 \%$ uncertainty in the determination of the AOM thermal lens (dashed and dotted curves). We ascribe this mismatch to some degree of astigmatism that affected the trapping beam for this specific $\delta z$ range, likely caused by a non-perfect centering of the beam on the AOM crystal. These non-ideal conditions enhance the thermal lensing effect since astigmatism significantly modifies the potential landscape experienced by the cold atomic cloud, yielding weaker effective confinement along the axial direction and amplifying the thermal shift of the trap minimum. On the other hand, we find quantitative agreement between the experimental data and the simulated curve around the $\Delta z_{th}$ zero crossing points, whose identification represents the main focus of our study. Most importantly, Fig. \ref{Fig5}c data confirm again the possibility to cancel out thermal lensing effects from a high power optical trapping setup by properly adjusting the AOM position with respect to the beam waist.

\section{Conclusions}

In conclusion, we have characterized the sources of thermal lensing associated with the various elements composing a typical high power setup for optical trapping of cold atomic clouds. From this survey, we identified the $TeO_2$ crystal of the AOM as the sole relevant thermal lens affecting the optical system, whereas we found that inexpensive fused silica lenses and quartz windows provide a negligible contribution.
We then devised a simple, totally passive scheme that enables to cancel thermal lensing effects on the trapping beam up to very high intensities. Our strategy relies on placing the thermal lens within one focus of the laser beam.
This allowed to stabilize the waist position of the high power beam used as optical dipole trap, with thermal shifts below our experimental resolution, as low as a few tens of microns. Our data are reasonably reproduced by a simple Gaussian beam matrices calculation, by treating the AOM crystal as a thin thermal lens $f_{th}$, employing the power dependence previously reported in literature for $TeO_2$ substrates \cite{Bogan_2015}.
Although this study was specifically oriented to the implementation of a high power optical dipole trap for cold atom experiments, our strategy may find applications within any generic optical setup featuring one or few thermal lensing sources.
Furthermore, this configuration could be also integrated into more complex setups, aiming to cure, besides thermal shifts of the focus position, thermal induced phase aberrations which can significantly distort the beam waist when this approaches the diffraction limit.

\section*{Funding}
This work was supported under European Research Council grant No. 637738 PoLiChroM, and under Italian MIUR FARE grant No. R168HMHFYM P-HeLiCS.

\section*{Acknowledgments}
We acknowledge insightful discussions with the members of the LENS Quantum Gases group, and in particular with Giacomo Roati and Francesco Scazza.


\bibliography{ThermalLensingBibliography}

\end{document}